**Low-temperature magnetic-field-driven thermal oscillator based on metal-superconductor joint**

Poonam Rani[1], Yoshikazu Mizuguchi[1]*

[1] Department of Physics, Tokyo Metropolitan University, Hachioji 192-0397, Japan.

* Corresponding author: Yoshikazu Mizuguchi (mizugu@tmu.ac.jp)

**Abstract**

Thermal control is one of the important technologies for fundamental science and thermal management. Among them, thermal oscillators have been in demands in the field of materials science and device application. In general, flexible frequency, amplitude, and waveform are needed for useful thermal oscillator, and the stability of the average temperature is also highly required. However, thermal oscillators based on an AC-current-driven heater require complicated control of input power to achieve the above-mentioned flexibility and stability of the outputs. Here, we demonstrate that magnetically-driven thermal oscillators fabricated using a metal-superconductor (Cu-Pb) joint achieve those requirements easily by tuning the applied magnetic field ($H$). A DC-current-driven heater is attached on the metal (Cu) side, and the superconductor (Pb) edge is attached to thermal bath. We use a sharp and huge change in thermal conductivity at the superconducting transition of the Pb wire to generate thermal oscillation at the Cu-wire side. A sine-shaped thermal oscillation with an amplitude of 180 mK and a frequency of 0.17 Hz is observed with highly stable average temperature. Furthermore, a larger amplitude is achieved in a square-shaped oscillation with a larger $H$ amplitude. Our thermal oscillator with temperature stability, large amplitude, and relatively high frequency will work as a flexible AC heat source at cryogenic temperatures.

**Highlights**

- Magnetically driven thermal oscillators based on a superconductor were fabricated.
- Generated waveform, frequency, amplitude, and average temperature are flexible.
- The obtained thermal oscillators will be useful for cryogenic sensor calibration.

**List of Abbreviations:**

PPMS - Physical Property Measurement System

TTO - Thermal Transport Option

MTS - Magneto-Thermal Switching

$T$ - Temperature

$T_c$ – superconducting transition

$T_{bath}$ - bath temperature

TES - Transition edge sensor

$t$ - time

AC - Alternating current

DC – Direct current

$H$ - Magnetic field

$H_c$ - Critical magnetic field

$\kappa$ - Thermal conductivity

$R$ – Thermal resistance

## 1. Introduction

Thermal control is one of the tools essential for development of thermal management technologies and for achieving precise measurements in fundamental sciences including quantum computing [1–6]. Previous studies have demonstrated heat flow manipulation using thermal diodes, switches, and phononic concepts, highlighting their importance in energy control and device applications. For example, sensors sensitive to thermal fluctuation should be calibrated or tested with thermal oscillation to certify and improve the performance. Precise measurements of magneto-caloric or thermoelectric effects and AC (alternating current) specific heat measurements also require precise thermal oscillation [7-9]. In addition, research on thermal logic circuit, which is a thermal analogue of electronic circuit, has been activated [3,10].

In parallel, significant efforts have been devoted to understanding thermal oscillations and amplitude modulation in fluid-based systems, including mixed convection flows, boundary layer dynamics, and nanofluid transport. Recent studies have reported wave-like thermal behavior, amplitude variation, and frequency characteristics in systems such as convection over cylinders, Darcy–Forchheimer nanofluid flows, chemically reactive and buoyancy-driven surfaces, and radiative magnetohydrodynamic flows under various physical conditions. These works highlight the importance of nonlinear thermal transport, coupling between heat transfer and flow dynamics, and the role of thermal conductivity in shaping oscillatory temperature profiles [11-15].

Despite these advances, such oscillatory thermal phenomena have been predominantly explored in fluid and continuum systems, where convection and flow-driven mechanisms govern heat transfer. In contrast, thermal oscillations in solid-state systems, particularly those involving superconducting materials, remain largely unexplored. To bridge this gap and enable new functionalities in solid-state thermal management, the development of controllable thermal oscillation techniques is highly desirable.

This paper reports on new method to create magnetically driven thermal oscillation with flexible amplitude, frequency, and waveshape using a bulk metal-superconductor joint. Superconducting materials have been explored for thermal switching due to their strong magnetic-field-dependent thermal conductivity [16–19]. As well known, electrons form Cooper pairs below the superconducting transition temperature ($T_c$), and superconducting gap opens. Therefore, at $T < T_c$, thermal conductivity ($\kappa$) becomes low due to the suppression of electronic contribution of $\kappa$. Using the suppression of $\kappa$, magneto-thermal switching (MTS) without mechanical motion can be achieved [16,17]. We demonstrated that the use of superconductors for MTS has merits of high MTS ratio [18] and nonvolatility of MTS with a relatively low applied field ($H$) [19]. Furthermore, superconductor-based thermal diodes have been fabricated using nano-scale junction and bulk materials, enabling rectification of heat flow [20–22]. Those results confirm the versatility of superconductors for thermal-control techniques. However, these studies mainly focus on static thermal control, and dynamic thermal oscillations in bulk solid-state systems remain largely unexplored. On thermal oscillation technique, Guarcello et al. reported thermal oscillation in a superconductor-based Josephson junction [23]. However, fabrication of a thermal oscillator based on bulk-scale superconductors has not been reported. The present work represents the first experimental demonstration of magnetically driven thermal oscillations in a bulk superconductor-based system, arising from the sharp transition in $\kappa$ of high purity Pb wire under an applied $H$. To the best of our knowledge, there are no prior studies have reported thermal oscillations based on such a sharp superconducting transition in bulk metal-superconductor system. Therefore, to expand the application of superconductor-based device for thermal control at cryogenic temperatures, new method to generate thermal oscillation with a temperature stability, large amplitude, and relatively high frequency using a bulk superconductor has been desired.

## 2. Experimental details

High-purity Pb (5N) and Cu (5N) wires with 0.5 mm diameter are the products of Nilaco corporation; the wires were taken from an evacuated desiccator just before the joint fabrication to avoid surface oxidation. Pb-Cu joint was made by conventional Sn-Pb solders. The typical joint sample size is about 5 cm in the total length, and the distance between $T_1$ and $T_2$ is about 4 cm, and that between $T_1$ and $T_3$ is about 1 cm (see Figs. 1, 5, 6 for definition). A Cu heater is attached near the edge of the Cu wire. Thermal measurements were performed on a Physical Property Measurement System (PPMS Dynacool, Quantum Design) using a sample puck for Thermal Transport Option (TTO). The heater and the Cernox thermometers were attached using Ag epoxy paste and Cu wire with a diameter of 0.2 mm. $\kappa$ of Pb and Cu wires were measured using the TTO. Magnetic field is applied to the direction parallel to the wire length for both the measurements on Pb wire, Cu wire, and the Pb-Cu joint sample.

## 3. Results

Figure 1a shows the thermal oscillation observed in the Pb60-Cu40 joint sample composed of Cu (normal-conducting) wire and a Pb (superconducting) wire. First, we explain the setup of the fabricated thermal oscillator and the measurement configuration. The thermal oscillator is designed as a thermal analog of the relaxation oscillator, which is a type of electronic circuit that produces various oscillations including square and triangular shapes [24]. As shown in Fig. 1b, the Cu and Pb wires with a diameter of 0.5 mm are soldered using Sn-Pb solder. On the Cu side, a heater and a Cernox thermometer ($T_1$) are attached, and the Pb side is mechanically fixed to the thermal bath ($T_{bath}$) of the sample puck whose temperature is variable and precisely controllable. Figures 1c and 1d explain the driving mechanism of the thermal oscillator. The heater is DC-current-driven, and the heater power is always constant during one measurement.

Figures 2(a) and 2(b) show the $\kappa$-$H$ for the used Pb and Cu wires at $T$ = 3 K, 4 K, 5 K and 6 K. Although Pb shows a superconducting transition accompanied by the huge change in $\kappa$, Cu does not show a superconducting transition and large magnetoresistance. We use the sharp change in $\kappa$ of Pb by tuning the applied $H$. By applying AC magnetic field (AC-$H$) at the limited $H$ range in superconducting transition in $\kappa$-$H$ (green-colored region in Fig. 1d), $\kappa$ of Pb wire oscillates depending on the change in number of Cooper pairs. The sharp change in $\kappa$ of Pb well tracks the frequency of AC-$H$, which is achieved by the characteristic thermal-transport property of high-purity Pb wire with $H$ parallel to the wire length direction [20]. Due to the change in $\kappa$, the rate of heat evacuation from the Cu side to the thermal bath also oscillates. Finally, $T_1$ oscillation with a waveshape similar to AC-$H$ is generated. Figure 1a is an example of sine-shaped oscillation generated by AC-$H$ (~50 Oe amplitude) shown in the inset of Fig. 1b. The average $T_1$ is 5.44 K, and the obtained data are nicely fitted by a sine function with a frequency of 0.167 Hz. Using the same strategy, triangular thermal waves can also be generated, as shown in Fig. S3. This demonstrates that the temperature response can be precisely designed by the applied magnetic field, reflecting the strong coupling between magnetic field and thermal transport.

Next, we increase $H$ amplitude to generate square-shaped wave and to investigate the performance of the thermal oscillator by changing the heater power and $T_{\mathrm{bath}}$. As shown in Fig. 3a, applying larger AC-$H$ amplitude results in the saturation of $\kappa$ change of Pb because the large change in $\kappa$ is limited to the green-arrow range in Fig. 1d. Therefore, the resulting $T_1$ oscillation is close to a square-shaped wave. Figure 3b shows an example obtained under a larger AC-$H$ amplitude, enabling a systematic comparison of the resulting oscillation amplitude as a function of heater power and $T_{\mathrm{bath}}$. The oscillatory thermal response was measured under an applied AC-$H$ within a specific field window near the $T_c$. In Fig. 3(b), $H$ was varied between 350 and 500 Oe, corresponding to a mean field of approximately 425 Oe with an amplitude of ~±75 Oe. As the heater power increases, the average sample temperature rises, thereby shifting the $T_c$ region. To sustain the oscillatory response, the $H$ window was correspondingly adjusted to higher values to match the elevated temperature range. For example, when the heater power is increased from 0.8 mW to 1.2 mW, the field range is shifted accordingly (from 350 and 500 Oe to 300 and 470 Oe) to ensure operation remains within the transition regime, where the $\kappa$ exhibits strong magnetic-field dependence and enables pronounced thermal oscillations. Similarly, measurements were carried out at different $T_{bath}$ to systematically investigate the temperature dependence of the thermal oscillations. Figure 3c shows the $T_1$ oscillation obtained at $T_{bath}$ = 5 K and heater power of 0.1–1.2 mW. It is clear that the $T_1$ amplitude is correlating with the heater power. The heater power dependence of $T_1$ amplitude is shown in Fig. 4a. In Fig. 3d, data taken with the heater power of 0.8 mW and at different $T_{bath}$ = 3–6 K are shown. The $T_1$ amplitude is not sensitive to the change in $T_{bath}$, which is a merit of use at variable temperatures below $T_c$ of the superconductor. As shown in Fig. 4b, the maximum amplitude is observed at $T_{bath}$ = 5 K, and the amplitude at $T_{bath}$ = 6 K is relatively lower.

A simplified thermal model was employed to estimate the $T_1$ under magnetic field modulation at a constant power of 0.8 mW. The calculated temperature variation, based on the field-dependent thermal conductance of the Pb and Cu segments (Table S1, Supplementary Material), is presented. Because of the difficulty of the inclusion of thermal resistance at the solder joint, we ignore that in this simplified model. To evaluate the validity of the thermal model, the calculated $T_1$ values are compared with experimentally obtained temperatures under identical conditions, as summarized in Table S1. The calculated $T_1$ values show reasonable agreement with the experimental results, with deviations attributed to the simplified model that does not include the thermal resistance, $R$ of the solder joint effects. This agreement indicates that the field-dependent $R$ of the Pb segment plays the dominant role, while the Cu wire, due to its relatively large heat capacity, contributes to the dynamic response and can be regarded as an effective thermal capacitance. To further validate and extend this analysis, more detailed numerical simulations (e.g., using COMSOL Multiphysics) will be carried out in future work.

In Fig. 5, we show the data obtained for a different Pb60-Cu40 sample (sample #2) for confirmation of the reproducibility. Similar thermal oscillation was observed with the same measurement conditions (Figure 5b). Furthermore, to confirm that the observed $T_1$ oscillation is not caused by the Eddy current due to AC-$H$, we display the whole data of $T_1$ and $T_2$, attached near the thermal bath, in Fig.5a. Here, the heater is tuned on during the first half of the measurement up to $t$ = 50s. Then, the heater is turned off, and the temperature difference between $T_1$ and $T_2$ becomes quite small. AC-$H$ is uniformly applied during the measurement (Figure 5c). The fact that $T_2$ does not show a clear oscillation under AC-$H$ confirms that the observed $T_1$ oscillation is caused by the mechanism proposed in this work (Figure 5d). Furthermore, after turning off the heater power at $t$ = 50 s, the $T_1$ oscillation is suddenly suppressed (Figure 5d), which also confirms that our assumption on the mechanism is correct.

## 4. **Discussion**

First, we briefly explain the flexibility of the material design. Since we just need a thermal capacitor, which should be a non-superconducting metal like Cu, and an MTS material like superconducting Pb, the possible combination is quite flexible. In addition, as displayed in Fig. S1, changing the ratio of Pb-Cu wire does not give a critical difference between Pb60-Cu40 and Pb50-Cu50. Therefore, the shape of the thermal oscillator can be tuned. The, we examine the uniformity of the generated thermal oscillation at the Cu wire by measuring $T_3$ on the Cu side near the solder joint (Fig. 6a). In Fig. S2, we show the time dependence of $T_1$ and $T_3$ for sample #3 and an example picture of the sample setup. As shown in Fig. 6b ($T_{bath}$ dependence) and Fig. S2d (heater power dependence), $T_1$ and $T_3$ exhibit almost the same thermal oscillation. Because the distance between $T_1$ and $T_3$ is about 10 mm, the spatially uniform thermal oscillation generated in the Cu wire should be achieved by the high $\kappa$ of Cu and its field-insensitivity. In contrast, the Pb wire side should have large fluctuation of oscillation because of oscillating $\kappa$ and the time-dependent average temperature under AC-$H$. Therefore, when considering practical application of thermal oscillators based on superconductor-metal joint, the presence of thermal capacitor (Cu wire in the present case) is highly important to produce uniform thermal oscillation with required surface size.

The merit of our thermal oscillator is that the oscillation control is done by the applied AC magnetic field. In the case of an AC heating driven by AC current of the heater, precise PID control is needed to obtain a desired oscillation form because the relaxation rates of Joule heating and cooling by the thermal bath are different. In the case of our thermal oscillator, however, the driving mechanism is basically common between heating and cooling because the change in $\kappa$ of Pb is sharp and exhibits nice tracking to the AC-$H$, which is based on the sensitivity of superconducting transition to the applied AC-$H$.

Now, we discuss potential application of this thermal oscillator. The applied AC-$H$ controls the thermal oscillations through its strong influence on $\kappa$ of the Pb segment. As the field varies across the $T_c$, $\kappa$ of Pb changes significantly between the superconducting (low $\kappa$) and normal (high $\kappa$) states, leading to modulation of heat flow. The amplitude of the temperature oscillations is determined by the magnitude of this conductivity change, with larger variation across the transition resulting in higher amplitude. If the field variation does not fully traverse this transition region, the amplitude is reduced. The oscillation frequency follows the applied AC-$H$, while at higher frequencies through high rate of change of $H$, the temperature response becomes smaller and delayed due to the finite thermal response time of the system. Furthermore, the waveform of the thermal oscillations reflects the temporal profile of the applied AC-$H$, with additional distortion arising from the nonlinear dependence of $\kappa$ on the $H$ near the $T_c$. The largest $T_1$ amplitude obtained in this work is 360 mK with 1.2 mW, but the amplitude can be further enhanced by tuning the heater power. The $T_1$ oscillation frequency was about 0.15 Hz in Fig. 3c. However, in this study, $H$ is controlled by rapid change in DC-driven superconducting magnet as described in Materials and Methods. Therefore, we expect that the maximum frequency, where the $T_1$ oscillation can track the AC-$H$ frequency, will be higher than the current data. In Fig. 3b, two datapoints are highlighted with red circles. Although we have not performed measurements on trackable frequency, these two points show 200 mK change in 0.77 s, which is corresponding to1.3 Hz. Therefore, if perfect AC field can be applied to the Pb wire, our thermal oscillator will work at frequencies higher than 1 Hz. From these potential conditions, we propose that our thermal oscillator can be useful in AC measurements at low temperatures, which includes specific heat and thermoelectric properties. In addition, the stable thermal oscillation will be useful for testing sensitivity of low-temperature sensors and correcting their precision. Furthermore, the frequency range will be useful for dynamic calibration sources for high-sensitivity cryogenic detectors including astro-science transition edge sensor (TES) [25, 26], while demonstration of the use of thermal oscillation for the calibration is needed. To explore the utility of the newly developed thermal oscillator based on bulk superconductor, further investigation of the performance and the driving mechanism and optimization of the device configuration is needed.

## 5. Summary

Thermal oscillators based on a superconductor-metal (Pb-Cu) joint have been proposed. Using the steep change in $\kappa$ at the superconducting transition ($T_c$ and $H_c$) in a Pb wire, thermal oscillation at the Cu side is achieved by applying AC magnetic field. A DC-current-driven heater is attached on the Cu side, and the Pb edge is attached to thermal bath. A sine-shaped thermal oscillation with an amplitude of 180 mK and a frequency of 0.17 Hz is observed with highly stable average temperature. A larger amplitude is achieved in a square-shaped oscillation with a larger $H$ amplitude, and the potential trackable frequency of thermal oscillation is higher than 1.3 Hz. We confirmed flexibility of the oscillator design through investigating the affection of the Pb-Cu ratio and the uniformity of the generated thermal oscillation at the different points of the Cu wire. Our thermal oscillator with temperature stability, large amplitude, and relatively high frequency will work as a flexible AC heat source at cryogenic temperatures.

**Acknowledgements**

The authors thank Y. Sakuraba, K. Uchida, Y. Ezoe, Y. Watanabe, A. Yamashita, and Y. Hattori for fruitful discussion. This work was partly supported by JST-ERATO (No.: JPMJER2201) and TMU research fund for young scientists.

**Data Availability**

All the data presented in this article can be provided by reasonable requests to the corresponding author.

**Author Contributions**

**Poonam Rani:** Conceptualization, Methodology, Software, Formal analysis, Investigation, Data Curation, Writing - Original Draft, Writing - Review & Editing, Visualization

**Yoshikazu Mizuguchi:** Conceptualization, Methodology, Validation, Formal analysis, Investigation, Resources, Data Curation, Writing - Original Draft, Writing - Review & Editing, Visualization, Supervision, Project administration, Funding acquisition

**Figures**

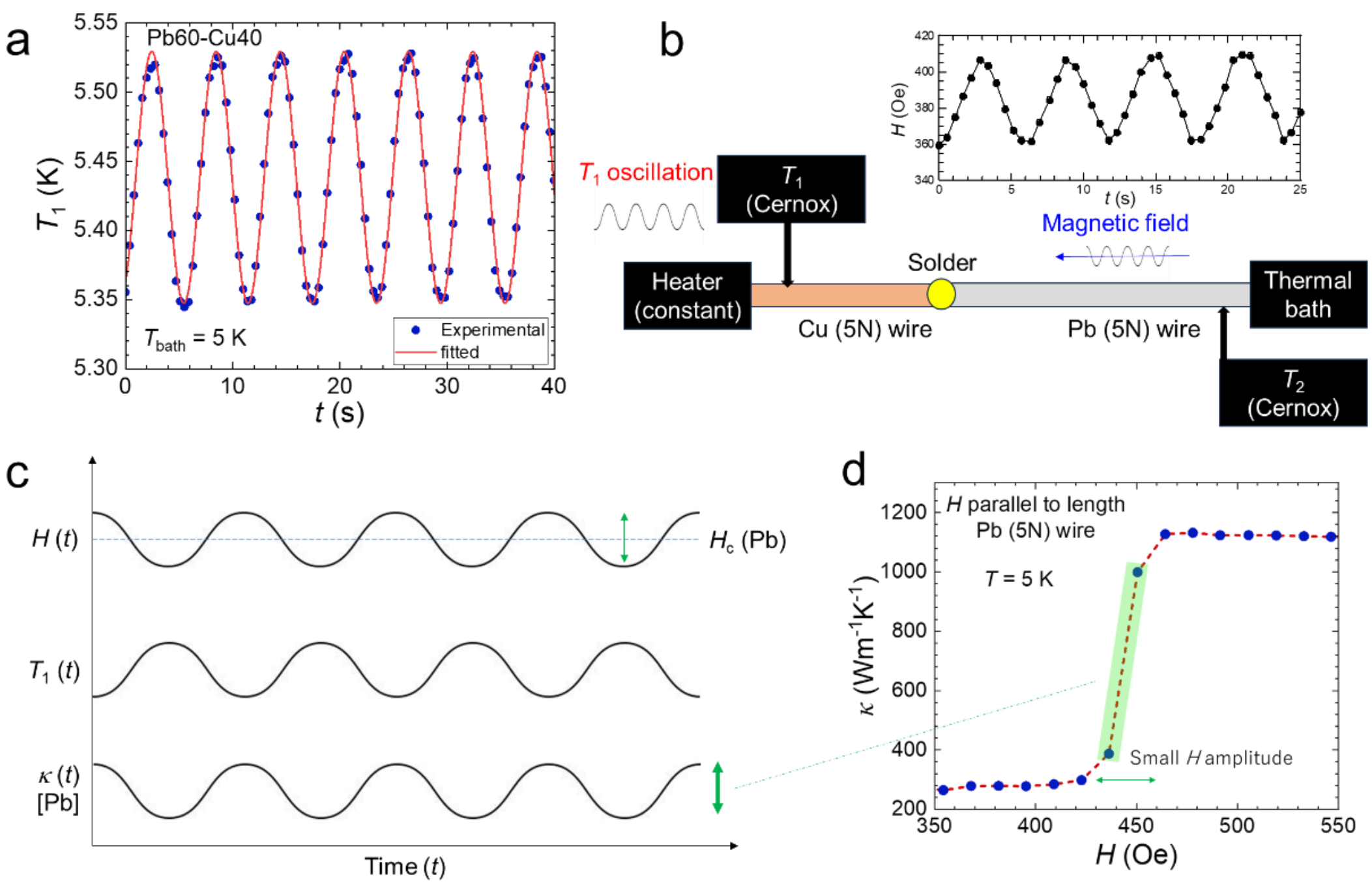

**Fig. 1. Thermal oscillation with a sinusoidal wave.** (a) Thermal ($T_1$) oscillation for the Pb60-Cu40 joint sample. The red curve is the fitting result with a sine function with 0.167 Hz. (b) Measurement setup and the applied magnetic field ($H$). (c) Schematic images of the time dependence of $H$, $T_1$, and thermal conductivity of Pb ($\kappa$). (d) $\kappa$-$H$ of a Pb wire measured at $T$ = 5 K. As indicated with green arrow, a huge change in $\kappa$ is achieved by a small difference in $H$.

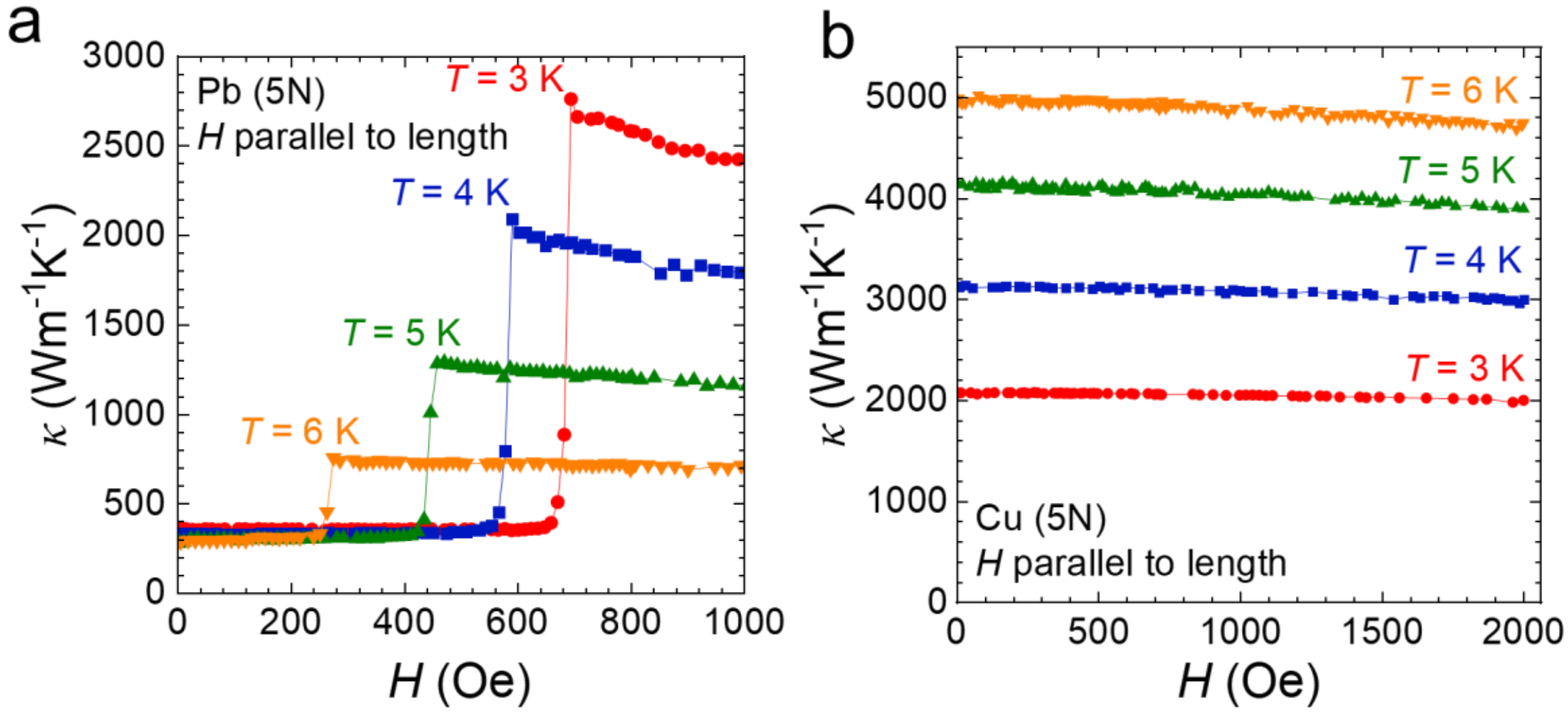

**Fig. 2. Thermal transport properties of Pb and Cu under magnetic fields.** (a,b) *κ*-*H* for high-purity (5N) wires of (a) Pb and (b) Cu at $T = 3$ K, 4 K, 5 K and 6 K. *H* is applied to the direction parallel to the wire length.

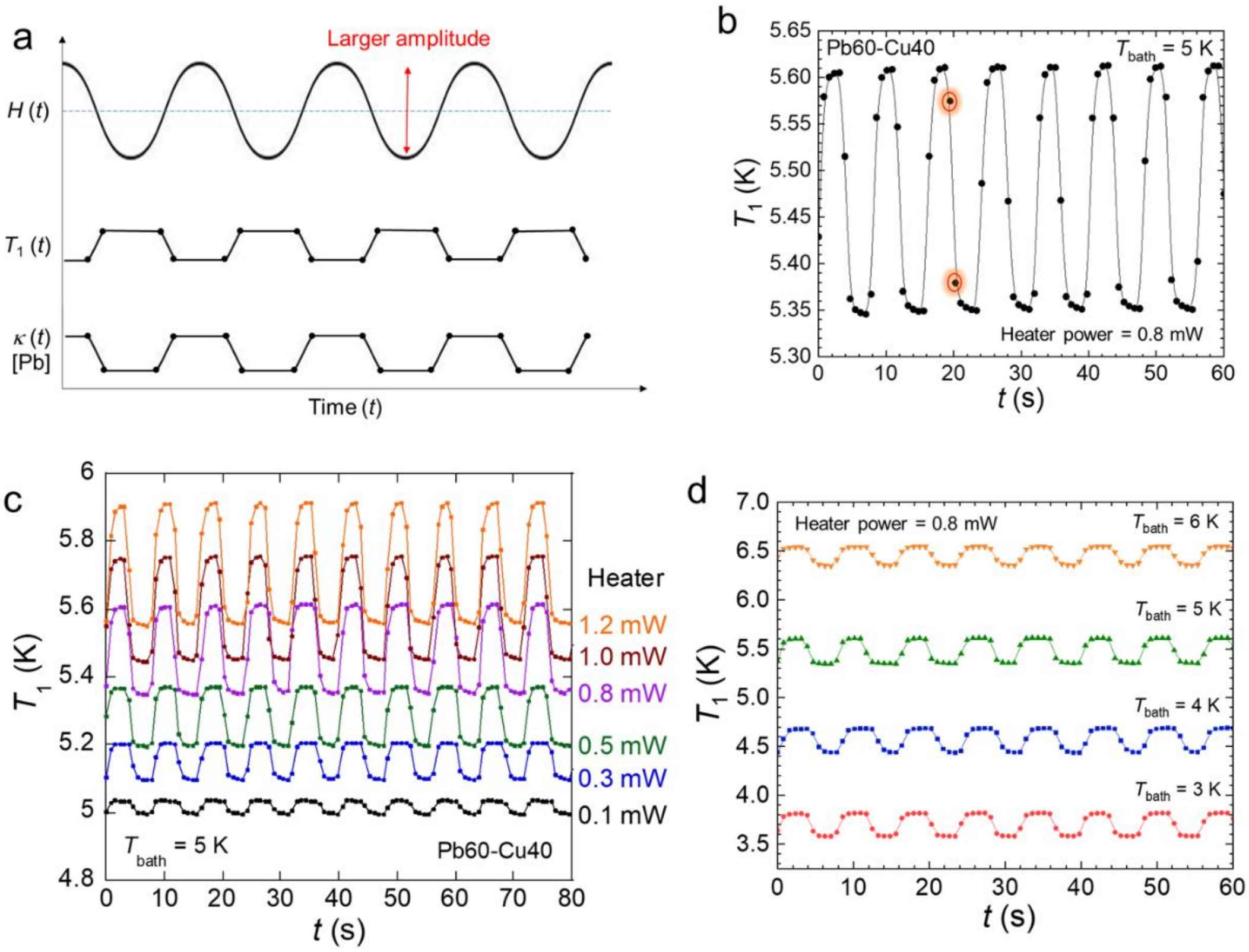

**Fig. 3. Heater power and $T_{bath}$ dependence of the generated oscillation.** (a) Schematic images of the time dependence of $H$, $T_1$, and $\kappa$ with a larger $H$ amplitude. (b) $T_1$ oscillation taken at $T_{bath}$ = 5 K with a heater power of 0.8 mW when AC magnetic field was applied in the range of 350–500 Oe, corresponding to a mean field of ~425 Oe with an amplitude of ~±75 Oe. Two red circles indicate typical datapoints showing a sharp change of $T_1$ (~200 mK) in a time interval of 0.77 s. (c,d) Comparison of $T_1$ oscillations taken with different (c) heater power and (d) $T_{bath}$.

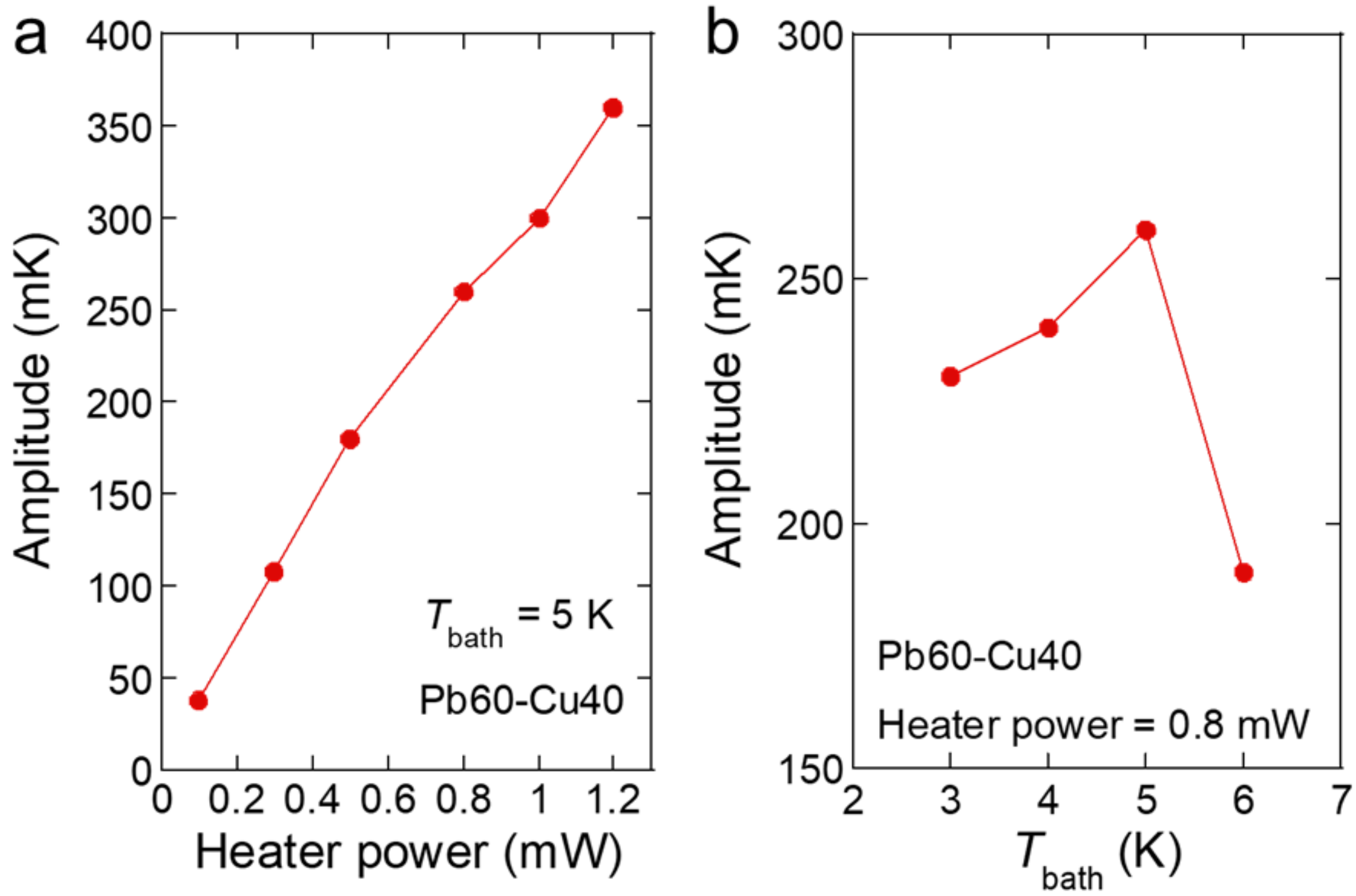

**Fig. 4. Comparison of thermal amplitude.** (a) Heater power dependence and (b) $T_{bath}$ dependence of the obtained $T_1$ amplitude for Pb60-Cu40. The data used for estimating amplitude are taken from Fig. 2.

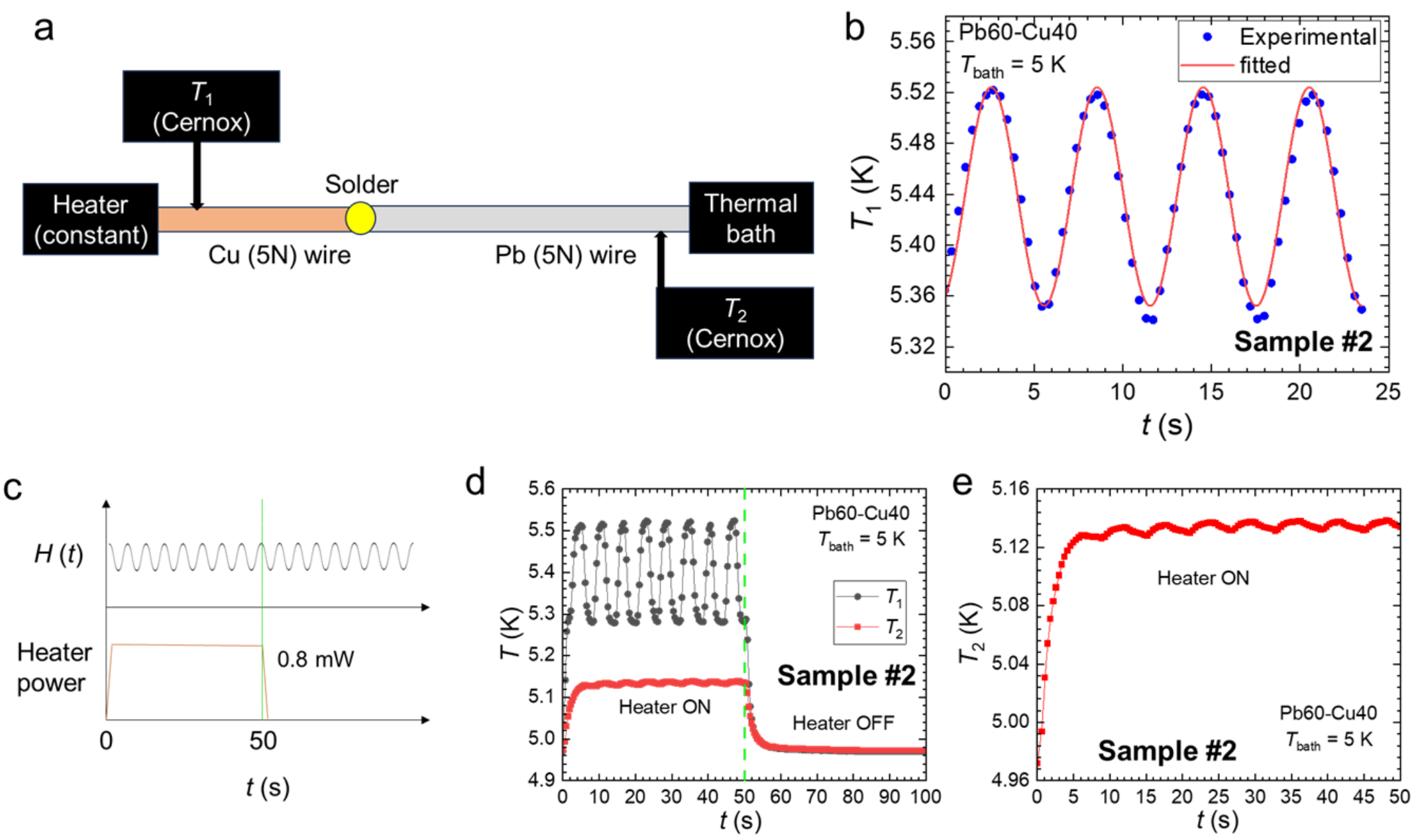

**Fig. 5. Reproducibility and temperature ($T_2$) scan during the measurement.** (a) Setup with $T_2$ at the lower-temperature Pb side. (b) Sine wave obtained for sample #2 (Pb60-Cu40). (c) Measurement sequence of heater power and applied field ($H$). (d, e) Time dependence of $T_1$ and $T_2$.

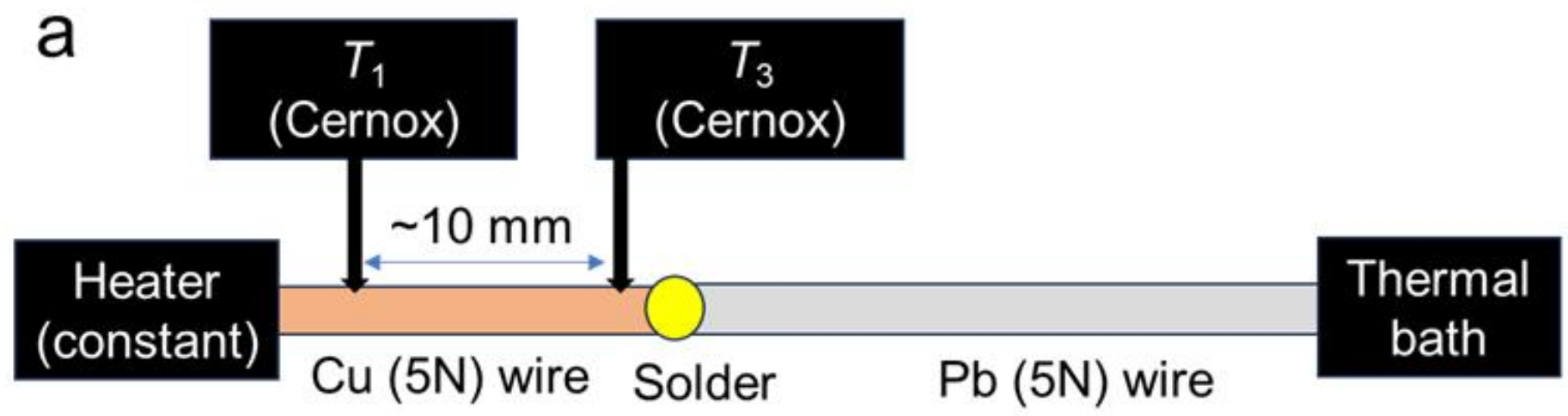

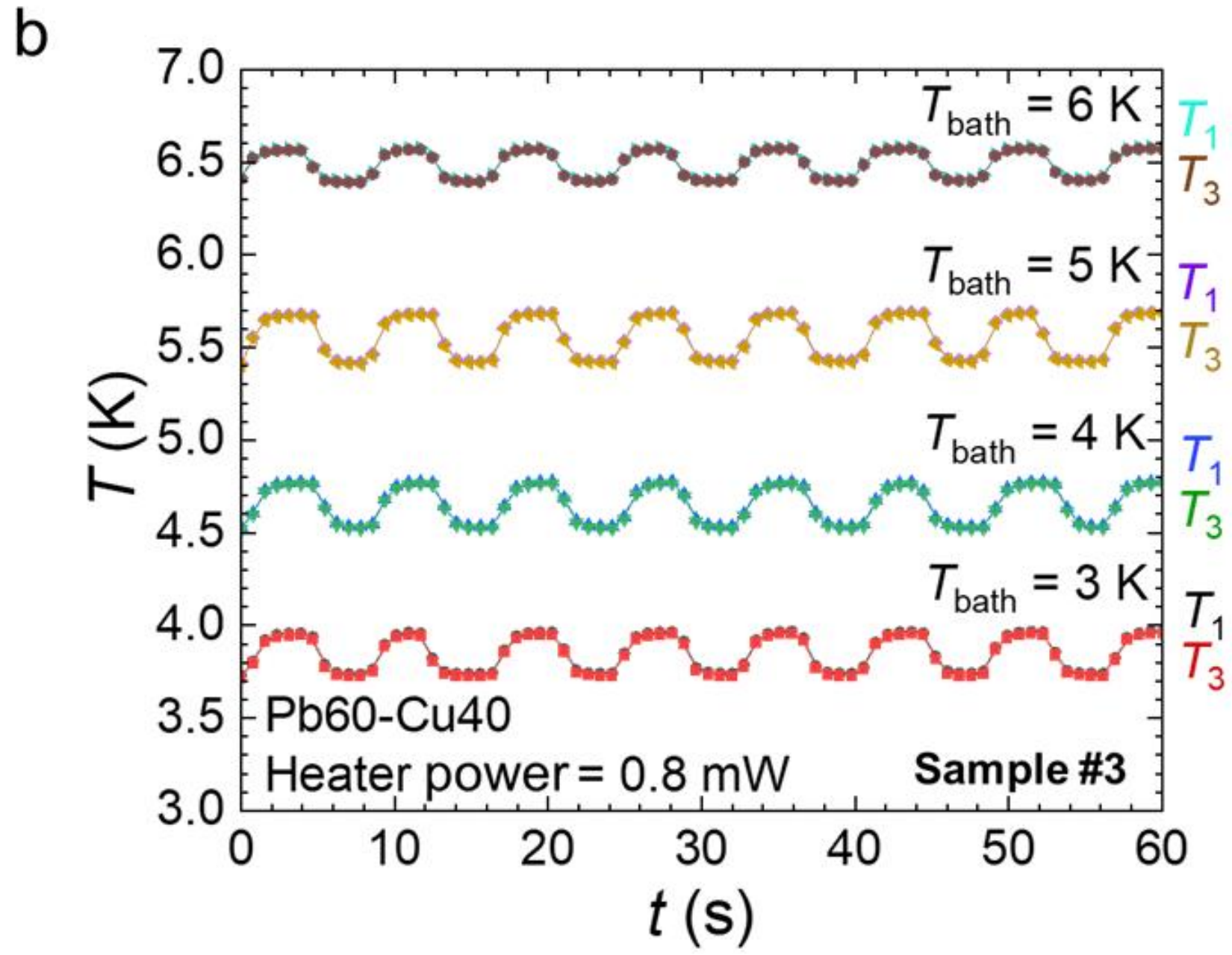

**Fig. 6. Uniformity of the generated oscillation at different positions on the Cu wire.** (a) Setup with $T_3$ at the Cu wire near the joint. (b) $T_{bath}$ dependence of the obtained $T_1$ and $T_3$ oscillations.

# Supplementary materials

**Low-temperature magnetic-field-driven thermal oscillator based on metal-superconductor joint**

Poonam Rani[1], Yoshikazu Mizuguchi[1]*

[1] Department of Physics, Tokyo Metropolitan University, Hachioji 192-0397, Japan.

* Corresponding author: Yoshikazu Mizuguchi (mizugu@tmu.ac.jp)

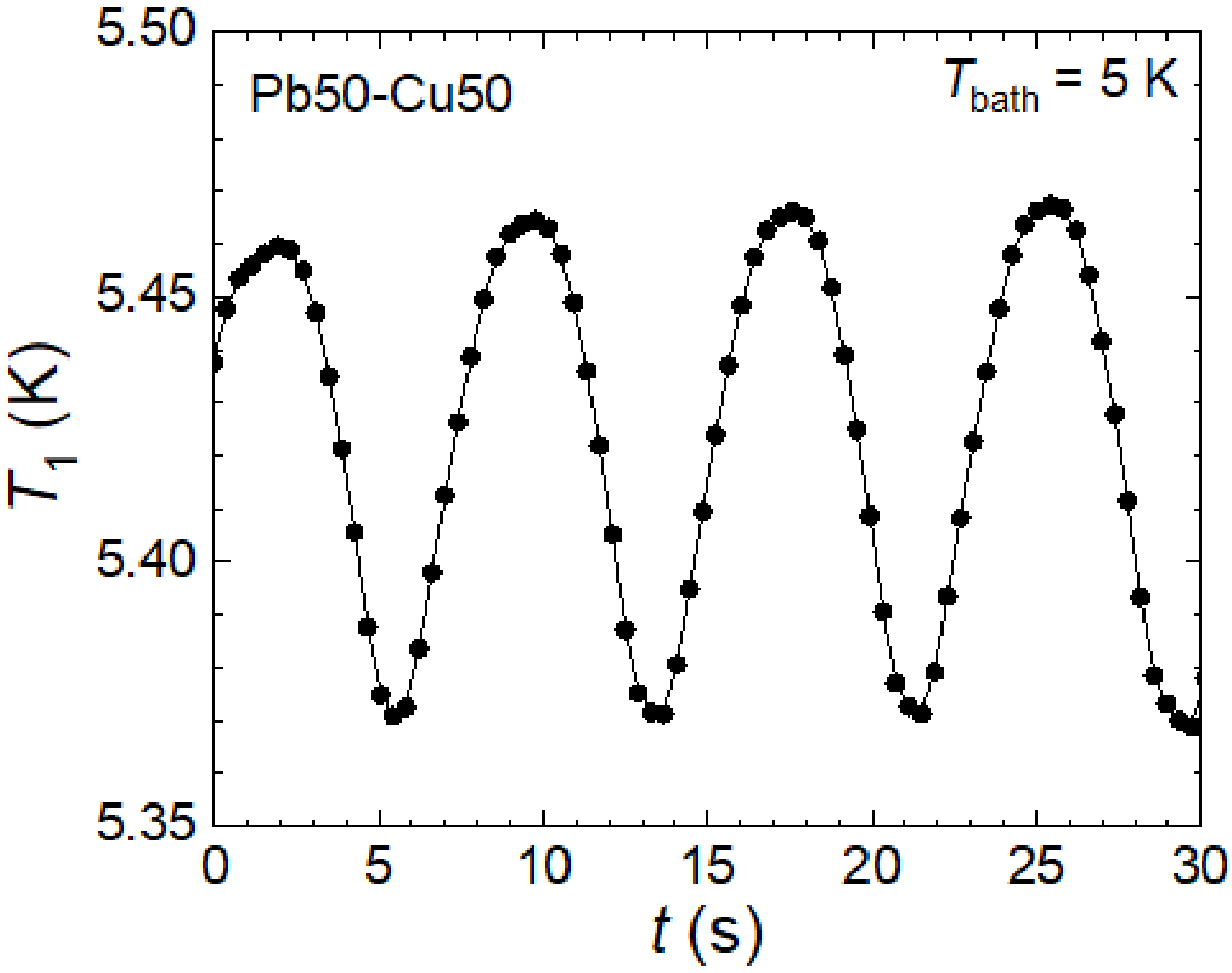

Fig. S1. Thermal oscillation obtained for the Pb50-Cu50 joint sample measured at $T_{bath}$ = 5 K and heater power of 0.8 mW.

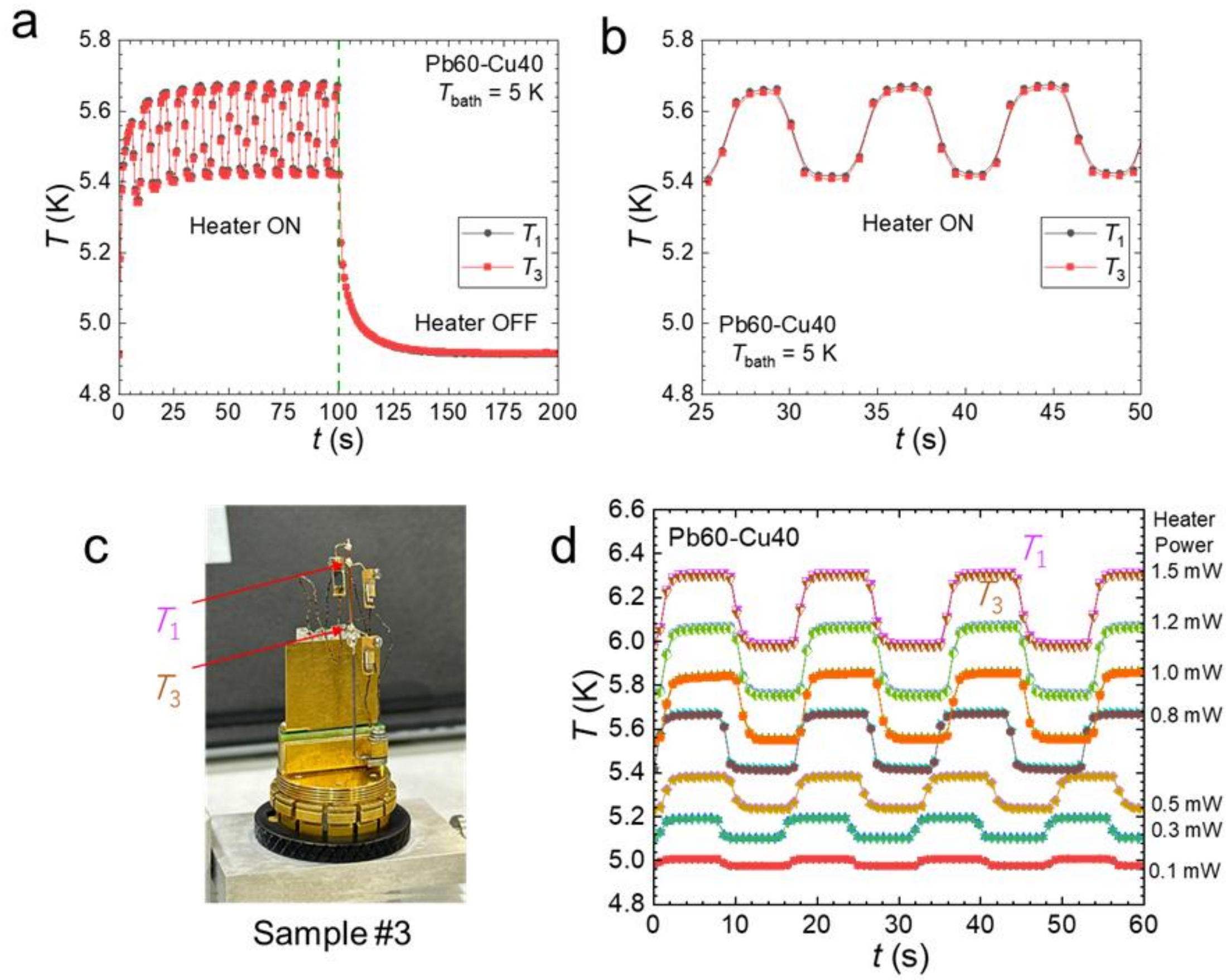

Fig. S2. (a,b) Time dependence of $T_1$ and $T_3$ oscillation with an oscillating magnetic field. (c) Picture of the sample measured with $T_1$ and $T_3$. (d) Heater power dependence of $T_1$ and $T_3$ oscillations.

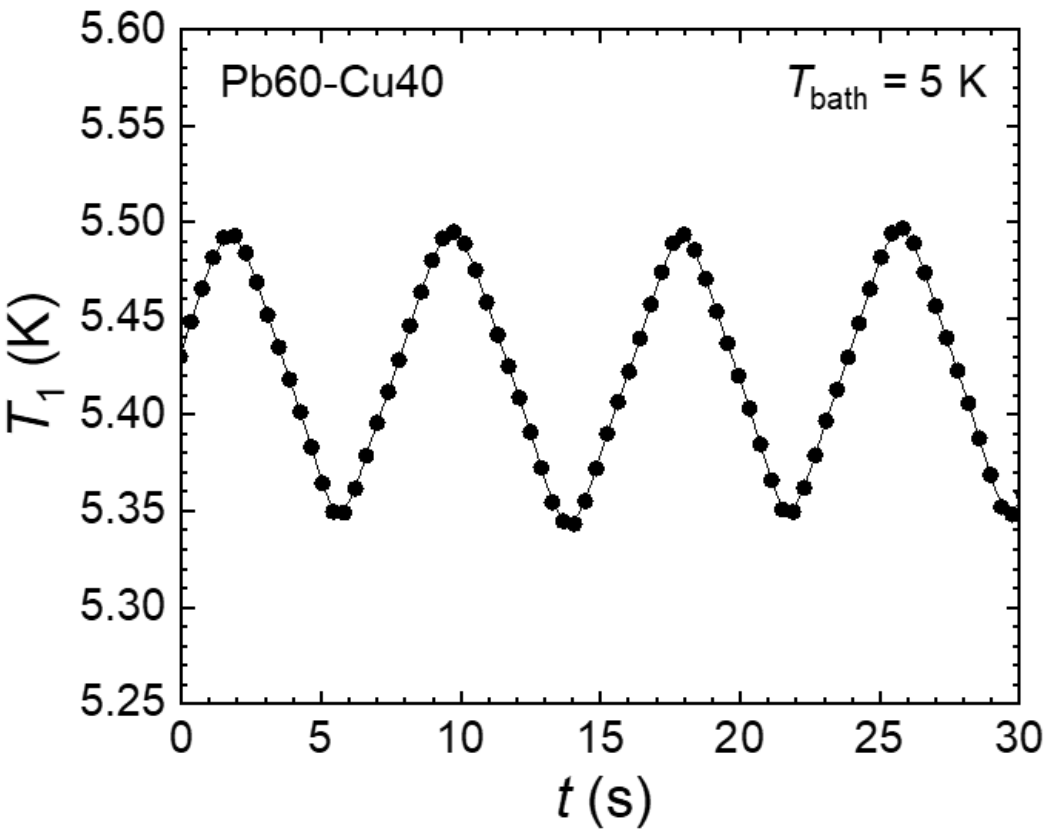

Fig. S3 Triangle-shaped thermal oscillation obtained for the Pb60-Cu40 joint sample measured at $T_{bath}$ = 5 K and heater power of 0.8 mW.

Table S1: Calculation of $T_1$ using $\kappa$ of Pb(5N) and Cu(5N) at different $T_{bath}$ when $H$ is applied within a particular range. Total $\kappa$ was calculated by simple summation of thermal resistance of wires.

| **$T_{bath}$ (K)** | ***H* (Oe)** | **$\kappa$_Pb (W/mK)** | **$\kappa$_Cu (W/mK)** | ***R*_Pb (K/W)** | ***R*_Cu (K/W)** | **Total *R* (K/W)** | **Total $\kappa$ (W/mK)** | **$T_1$ (K) (Calc.)** | **$T_1$ (K) (Exp.)** |
|---|---|---|---|---|---|---|---|---|---|
| 6 | 300 | 687 | 4983 | 177.27 | 17.04 | 444.30 | 408.28 | 6.41 | 6.35 |
| 6 | 200 | 286 | 5003 | 425.87 | 16.97 | 692.78 | 261.84 | 6.64 | 6.54 |
| 5 | 500 | 1257 | 4115 | 96.88 | 20.63 | 117.51 | 1543.66 | 5.11 | 5.34 |
| 5 | 350 | 305 | 4094 | 399.29 | 20.74 | 420.02 | 431.88 | 5.39 | 5.53 |
| 4 | 630 | 1988 | 3125 | 61.26 | 27.17 | 88.42 | 2051.48 | 4.08 | 4.44 |
| 4 | 500 | 341 | 3115 | 357.14 | 27.25 | 384.39 | 471.92 | 4.35 | 4.68 |
| 3 | 730 | 2648 | 2061 | 45.99 | 41.19 | 87.18 | 2080.76 | 3.08 | 3.58 |
| 3 | 600 | 353 | 2064 | 344.99 | 41.13 | 386.13 | 469.79 | 3.36 | 3.82 |